\documentclass[12pt]{iopart}

\usepackage{iopams}
\usepackage[T1]{fontenc}
\usepackage[utf8]{inputenc}
\usepackage{float}
\usepackage{graphicx}
\usepackage{amssymb}

\expandafter\let\csname equation*\endcsname\relax

\expandafter\let\csname endequation*\endcsname\relax

\usepackage{amsmath}
\usepackage{enumerate}

\makeatletter

\newcommand*{\diff}{\mathop{\!\mathrm{d}\!}}
\newcommand*{\sinc}{\mathop{\mathrm{sinc}}}
\newcount\hour \newcount\minute
\hour=\time \divide \hour by 60 \minute=\time \count99=\hour
\multiply \count99 by -60 \advance\minute by \count99 \hour=\time
\divide \hour by 60
\date{\today} 

\makeatother

\begin{document}

\title[Von Mises distribution and an infinite series ansatz for self-propelled particles]{Critical assessment of von Mises distribution and an infinite series ansatz for self-propelled particles}

\author{Rüdiger Kürsten and Thomas Ihle}
\address{Institut für Physik, Ernst-Moritz-Arndt Universität Greifswald, \mbox{Felix-Hausdorff-Str. 6}, 17489 Greifswald, Germany}
\ead{ruediger.kuersten@uni-greifswald.de}

\begin{abstract}
	We consider a Vicsek model of self-propelled particles with bounded confidence, where each particle interacts only with neighbors that have a similar direction. Depending on parameters, the system exhibits a continuous or discontinuous polar phase transition from the isotropic phase to a phase with a preferred direction.
	In a recent paper \cite{LSD15} the von Mises distribution was proposed as an ansatz for polar ordering. 
	In the present system the time evolution of the angular distribution can be solved in Fourier space. 
	We compare the results of the Fourier analysis with the ones obtained by using the von Mises distribution ansatz. 
	In the latter case the qualitative behavior of the system is recovered correctly. However, quantitatively there are serious deviations.
	We introduce an extended von Mises distribution ansatz such that a second term takes care of the next two Fourier modes. With the extended ansatz we find much better quantitative agreement. As an alternative approach we also use a Gaussian and a geometric series ansatz in Fourier space. The geometric series ansatz is analytically handable but fails for very weak noise, the Gaussian ansatz yields better results but it is not analytically treatable.
\end{abstract}
\maketitle
\section{Introduction}

In recent years, the collective motion of self-propelled particles has been subject to intensive research \cite{EWG15, MJRLPRS13, VZ12, Ramaswamy10}.
Active particles under consideration could be living subjects like e.g. birds, fish, bacteria or artificial nano or micro objects such as bimetallic nanorods, Janus particles or rotating disks.
A popular and simple model for interacting active matter is the Vicsek model (VM) \cite{VCBCS95, CSV97, CV00} or variants of it.
The dynamics of the VM consists of two steps, streaming and collision.
In the streaming period point particles move uniformly, all at the same speed but along individual directions.
The collision occurs instantaneously after the streaming period.
There, each particles adopts the average direction of all neighboring particles with some random perturbation (noise).

For large noise intensities the direction of each particle is random and the systems steady state is homogeneous and in polar disorder.
If the noise intensity is decreased particles start to align and move in a preferred direction. The system obtains a polar order.
For large system sizes the transition appears discontinuous and the polar ordered state is inhomogeneous \cite{GC04, CGGR08}. There are soliton-like density waves.
However, for small system sizes the polar ordered state remains homogeneous and the transition appears continuous.
In that case the transition can be studied via considering solely the angular distribution of all particle directions.
Even for the homogeneous solutions the dynamics remains complex and usually further simplifications are achieved by assuming molecular chaos and neglecting collisions of more than two particles.
Even so the angular distribution has still infinitely many degrees of freedom that evolve in a nonlinear fashion. 
A reduction of the complexity of the dynamics to finite dimensions is necessary for an analytical treatment and therefore a description of the angular distribution by only a few degrees of freedom is desirable.

In this paper we investigate four methods of complexity reduction.
As an example system for polar ordering we choose the VM with bounded confidence interactions \cite{RLI14}.
Such interactions are used in social sciences for the description of opinion dynamics \cite{DNAW00, HK02, BS15} to mimic the tendency of agents to be influenced only by others that have a similar opinion.  
In the bounded confidence VM, for a given particle, another nearby particle is seen as a neighbor only if it has a similar direction that differs by no more than $\alpha$.
For $\alpha=\pi$ the standard VM is recovered but for smaller values of $\alpha$ the system behavior is richer. 
There is a critical value $\alpha_c$ below which the transition appears discontinuous even for homogeneous systems.
To compare the performance of different approaches of complexity reduction we find, as a point of reference, the stable fixed points of the system investigating the full dynamics of the first $200$ Fourier modes.

Investigating the angular distribution in Fourier space we consider the first $l$ modes exactly and assume that all higher modes decay like a geometric series, following \cite{RLI14}.
With this ansatz all infinitely many modes can be treated analytically.
We find quantitatively good agreement with the full system, except for very weak noise where the ansatz breaks down.

For small noise a sharply peaked angular distribution is expected.
As a result one finds a Gaussian decay of higher Fourier modes instead of a geometric series.
Such an ansatz was used before to describe the distribution of coupled phase oscillators \cite{ZNFS03}.
With the Gaussian ansatz we find good quantitative agreement with the full system over the full range of noise intensities.
However, we are not able to evaluate the time evolution equation analytically for all modes such that the series has to be truncated after finitely many modes.

An alternative approach is to make an ansatz directly for the angular distribution. 
The von Mises distribution was proposed to describe polar ordering in a recent paper \cite{LSD15}, it was also used before \cite{DFL13, CK14}.
It has only a single parameter, the polar order parameter and it has the largest entropy of all angular distributions with the same order parameter, see e.g.\cite{JS01}.
In the present system we find that the ansatz yields qualitative agreement with the original system, however, quantitatively there are serious deviations, even for the standard VM.

To improve this direct approach we extend the ansatz by adding another term to the von Mises distribution.
The extended von Mises ansatz yields quantitatively much better agreement.
Close to the critical point and for small noise strength there is very good accordance with the original system.
In between there are small deviations.

The paper is organized as follows.
In Sec.~\ref{sec:model} we introduce the model and give the time evolution equation of the angular distribution assuming homogeneous systems, molecular chaos and neglecting collisions of more than two particles.
In Sec.~\ref{sec:fourier} we derive explicitly the time evolution equation of all Fourier modes of the angular distribution.
In Sec.~\ref{sec:seriesansatz} we test the performance of the geometric and Gaussian series ansatz in Fourier space.
In Sec.~\ref{sec:vonmises} we study the von Mises distribution ansatz and in Sec.~\ref{sec:extendedvonmises} an extension of it.
In Sec.~\ref{sec:conclusions} we summarize and compare the advantages and disadvantages of the four tested approaches for the description of polar ordering in homogeneous active systems.
Several technical derivations have been moved to the appendix.

\section{Model\label{sec:model}}
We investigate the two-dimensional Vicsek model with bounded confidence interactions that was introduced and analyzed in \cite{RLI14}.
We consider $N$ point particles with positions $x_i(t)\in \mathbb{R}^{2}$ that all move at constant speed $v_0$ into directions $\theta_{i}(t)$.

The dynamics consists of two parts, streaming and collision, that are alternated.
In the streaming period of length $\tau$ all particles perform a ballistic motion.
Hence the particle positions change to
\begin{align}
	x_{i}(t+\tau) = x_{i}(t) + \tau v_0 \binom{\sin(\theta_i)}{\cos(\theta_i)}.
	\label{eq:streaming}
\end{align}

After the streaming period the particles instantaneously change their direction of motion due to collisions. Let $\tilde{\theta}_i$ denote the precollisional directions. Then the directions after collision are given by
\begin{align}
	\theta_i = \Phi_i + \xi_i, \qquad
	\Phi_i = \arctan \Bigg[ \Big( \sum_{j\in \{i\} } \sin \tilde{\theta}_j  \Big) \bigg/  \Big( \sum_{j\in \{i\} } \cos \tilde{\theta}_j  \Big)      \Bigg],
	\label{eq:collision2}
\end{align}
where $\xi_i$ are independent random variables that are drawn uniformly from the interval $[-\eta/2, \eta/2]$.
The set of neighbored particles of particle $i$ is denoted by $\{i\}$.
Here $j\in \{i\}$ if $|x_i-x_j|\le R$ and $|\theta_i - \theta_j|_{mod \pi}\le \alpha$.
That means each particle interacts with all particles that are no further away than $R$ and that have directions that differ by no more than $\alpha$.
This is the bounded confidence interaction that ignores particles with too different directions.

Note that all particles interact with themselves. 
That means $\{ i\}$ is never an empty set since $i \in \{ i \}$.
For $\alpha=\pi$ the standard Vicsek model is recovered.

Let $f(x, \theta, t)$ denote the density of particles that are at time $t$ located at position $x$ and move into direction $\theta$.
We assume molecular chaos that means we make the assumption that all particles are independent and identical distributed before the collision
\begin{align}
	p_{N}(x_1, \theta_1, \dots, x_N, \theta_N, t) = \prod_{i=1}^{N} p_{1}(x_i, \theta_i, t)
	\label{eq:molecularchaos1}
\end{align}
and hence
\begin{align}
	f(x,\theta, t) = N p_{1}(x, \theta, t),
	\label{eq:molecularchaos2}
\end{align}
where $p_{N}$ and $p_{1}$ denote the $N$-particle and one-particle probability density function, respectively.
Assume that the collision occurs at time $t$ and denote the density before the collision by $\tilde{f}(x, \tilde{\theta}, t)$ and after the collision by $f(x, \theta, t)$.
They are then related by
\begin{align}
	&f(\theta, x, t)= \sum_{k=1}^{N} \binom{N-1}{k-1}\int_{[0, 2\pi]^k} \diff \tilde{\theta}_1 \dots \tilde{\theta}_{k} \frac{1}{\eta}\int_{-\eta/2}^{\eta/2}\diff \xi 
	\notag
	\\
	&\times \hat{\delta}(\theta-\xi - \Phi_1(\tilde{\theta}_1, \dots, \tilde{\theta}_k))\tilde{f}(x , \tilde{\theta}_1, t) 
	 \Big(1-\frac{M}{N}\Big)^{N-k} \prod_{i=2}^{k}\int_{|x_{1}-x'_{i}|\le R} \diff x'_i \frac{\tilde{f}(x'_i, \tilde{\theta}_i, t)}{N},
	\label{eq:fulltimeevolution}
\end{align}
where $M$ depends on the interaction radius $R$ and on the position $x$
\begin{align}
	M(x,t) :=  \int_{|x-x'|\le R} \diff x' \int_{0}^{2\pi} \diff \theta f(x, \theta, t)
	\label{eq:particlesincells}
\end{align}
and $\Phi_1(\tilde{\theta}_1, \dots, \tilde{\theta}_k)$ is given by Eq.~\eqref{eq:collision2} assuming that particles $1$ to $k$ are the only neighbors of particle $1$.

Eq.~\eqref{eq:fulltimeevolution} can be understood as follows. 
Let the first particle be located at position $x$ and move in direction $\theta$. 
Then $f(x, \theta, t)/N$ denotes the first particles probability distribution after the collision.
The number of particles that are within interaction distance of the first particle are denoted by $k$.
Since the first particle is always in interaction distance with itself, $k$ is at least one. 
The combinatorial factor $\binom{N-1}{k-1}$ gives the number of possibilities to choose $k-1$ neighbors for the first particle different from itself out of the $N-1$ other particles. The last line gives the probability that the particles one to $k$ are the only particles within interaction distance of the first particle. The delta function in the second line incorporates the collision rule. The integral over $\xi$ gives the expectation value of the noise term and the integrals over $\tilde{\theta}_1, \dots, \tilde{\theta_k}$ give the expectation with respect of the precollisional orientation. The orientations of the particles outside the interaction region are irrelevant.

We consider only homogeneous (in $x$) solutions 
\begin{align}
	f(x, \theta, t) = f(\theta, t)
	\label{eq:homogeneaous}
\end{align}
that are normalized such that
\begin{align}
	\int_{0}^{2\pi} \diff \theta f(\theta, t) = \int_{0}^{2\pi}\diff \theta f(x, \theta, t) = \rho (x) = \rho_0.
	\label{eq:normf}
\end{align}
Hence $M(x,t)$ is a constant
\begin{align}
	M(x, t)=\pi R^2 \rho_0.
	\label{eq:density}
\end{align}
and it follows from Eq.~\eqref{eq:fulltimeevolution} that
\begin{align}
	&f(\theta, t+\tau)= \sum_{k=1}^{N} \binom{N-1}{k-1}\int_{[0, 2\pi]^k} \diff \tilde{\theta}_1 \dots \tilde{\theta}_{k} \frac{1}{\eta}\int_{-\eta/2}^{\eta/2}\diff \xi 
	\notag
	\\
	&\times \hat{\delta}(\theta-\xi - \Phi_1(\tilde{\theta}_1, \dots, \tilde{\theta}_k))f(\tilde{\theta}_1, t) 
	\prod_{i=2}^{k}  f(\tilde{\theta}_i, t) \Big(1-\frac{M}{N} \Big)^{N-k+1} \Big( \frac{M/\rho_0}{N} \Big)^{k-1},
	\label{eq:homogeneoustimeevolution}
\end{align}
where we used that $f(\theta)$ is not effected by streaming.

We consider low densities $M \ll 1$ such that three particle interactions are unlikely and we can abort the series in Eq.~\eqref{eq:homogeneoustimeevolution} after the second term.
However the truncation of the series introduces a small error that leads to a violation of conservation of the particle density.
This has to be corrected by normalizing $f(\theta, t)$ to $\rho_0$, see also supplemental material of Ref.~\cite{Ihle13}. Performing the thermodynamic limit $N \rightarrow \infty$ we obtain
\begin{align}
	&f(\theta, t+\tau) = \frac{1}{1+M} \int_{0}^{2\pi} \diff \tilde{\theta}_1 \frac{1}{\eta} \int_{-\eta/2}^{\eta/2} \diff \xi 
	\hat{\delta}(\theta - \xi - \tilde{\theta}_1) f(\tilde{\theta}_1, t)
	\notag
	\\
	&+ \frac{M/\rho_0}{1+M} \int_{0}^{2\pi} \diff \tilde{\theta}_1 \int_{0}^{2\pi} \diff \tilde{\theta}_2 \frac{1}{\eta} \int_{-\eta/2}^{\eta/2} \diff \xi 
	\hat{\delta}(\theta - \xi - \Phi_1(\tilde{\theta}_1, \tilde{\theta}_2)) f(\tilde{\theta}_1, t) f(\tilde{\theta}_2, t).
	\label{eq:lowdensitytimeevolution3}
\end{align}
It is reasonable to consider the angular probability distribution of a single particle
\begin{align}
	p(\theta,t) = \frac{1}{\rho_0} f(\theta,t)
	\label{eq:angularprob}
\end{align}
that is normalized to one.

The bounded confidence interaction rule with parameter $0\le \alpha \le \pi$ yields for two particles
\begin{align}
	\Phi_1(\tilde{\theta}_1, \tilde{\theta}_2) = \begin{cases}
		\frac{\tilde{\theta}_1+ \tilde{\theta}_2}{2} \text{ if } |\tilde{\theta}_1-\tilde{\theta}_2|\le \alpha,\\
		\frac{\tilde{\theta}_1+ \tilde{\theta}_2}{2}+\pi \text{ if } 0<|\tilde{\theta}_1-\tilde{\theta}_2| - 2\pi \le \alpha,\\
		\tilde{\theta}_{1} \text{ else}.
	\end{cases}
	\label{eq:interactionrule}
\end{align}
Inserting this expression and Eq.~\eqref{eq:angularprob} into the time evolution equation \eqref{eq:lowdensitytimeevolution3} one obtains
\begin{align}
	&p(\theta, t+\tau) = \frac{1}{1+M} \frac{1}{\eta} \int_{-\eta/2}^{\eta/2} \diff \xi p(\theta - \xi, t)
	\notag
	\\
	&+ \frac{M}{1+M} \bigg\{ \int_{-\alpha}^{\alpha} \diff \hat{\theta} \frac{1}{\eta} \int_{-\eta/2}^{\eta/2} \diff \xi 
	p(\theta - \xi -\frac{\hat{\theta}}{2}, t) p(\theta - \xi +\frac{\hat{\theta}}{2}, t)
	\notag
	\\
	&+ \bigg( \int_{-\pi}^{-\alpha} \diff \hat{\theta} +  \int_{\alpha}^{\pi} \diff \hat{\theta}  \bigg) 
	\frac{1}{\eta} \int_{-\eta/2}^{\eta/2} \diff \xi 
	p(\theta - \xi, t) p(\theta -\xi +\hat{\theta}, t)\bigg\}.
	\label{eq:lowdensitytimeevolution4}
\end{align}
From this, substituting $\theta-\xi$, we can calculate
\begin{align}
	&\int_{0}^{2\pi} \diff \theta \cos(k\theta)p(\theta, t+\tau)= \frac{2}{k\eta}\sin(k\eta/2)\frac{1}{M+1}
	\label{eq:expectationcosine}
	\\
	&\times\bigg\{ \int_{0}^{2\pi}\diff \theta \cos(k\theta)p(\theta)
	+ M\Big[ \int_{0}^{2\pi} \diff \theta \cos(k\theta)\int_{-\alpha}^{\alpha}\diff \hat{\theta} p(\theta-\frac{\hat{\theta}}{2},t)p(\theta + \frac{\hat{\theta}}{2}, t) 
	\notag
	\\
	&+ \int_{0}^{2\pi} \diff \theta \cos(k\theta)\int_{[-\pi, -\alpha] \cup [\alpha, \pi]}\diff \hat{\theta} p(\theta,t)p(\theta + \hat{\theta}, t)\Big]   \bigg\}.
	\notag
\end{align}

\section{Fourier analysis\label{sec:fourier}}
To further investigate the time evolution Eq.~\eqref{eq:lowdensitytimeevolution4} we study the Fourier modes at time $t$
\begin{align}
	p(\theta, t) &= \sum_{k=0}^{\infty} g_k(t)\cos(k\theta).
	\label{eq:fouriercomponents}
\end{align}
The first mode $g_1$ is directly related to the order parameter $\Psi$ defined by
\begin{align}
	\Psi:= \langle \cos(\theta)\rangle = \int_{0}^{2\pi}\diff \theta \cos(\theta) p(\theta) = \pi g_1.
	\label{eq:orderparamvsfirstmode}
\end{align}
If $\Psi=1$ all particles move in the same direction, there is perfect order. 
If on the other hand $g_k=0$ for all $k\ge 1$ there is no order at all and $\Psi=0$.
The time evolution Eq.~\eqref{eq:lowdensitytimeevolution4} yields for $k>0$
\begin{align}
	g_{k}(t+\tau)= \frac{\lambda_k}{1+M}\bigg[ \frac{1}{2} g_k + 2\pi M \sum_{p=0}^{\infty}\sum_{q=0}^{\infty} B_{kpq}(\alpha) g_p g_q   \bigg],
	\label{eq:timeevolutionfourier}
\end{align}
where all modes on the right hand side are considered at time $t$ and
\begin{align}
	\lambda_k= \frac{4}{k\eta}\sin(k\eta/2).
	\label{eq:lambdak}
\end{align}
Eqs.~\eqref{eq:timeevolutionfourier} and \eqref{eq:lambdak} were given in \cite{RLI14}, where only some of the coupling coefficients $B_{kpq}(\alpha)$ have been calculated. 
However we can calculate all $B_{kpq}(\alpha)$ explicitly, cf. Appendix \ref{app:coupling}, resulting in the time evolution equation
\begin{align}
	g_{k}(t+\tau) =& \frac{\lambda_k}{1+M} \bigg\{ g_k \Big[ \frac{1}{2} + g_{0} M \big\{ \pi-\alpha +\frac{3}{k}\sin(k\alpha/2) 
	- \frac{1}{2k}\sin(k\alpha)  \big\}     \Big]
	\notag
	\\
	&+\frac{M}{2}\alpha \sum_{q=1}^{\infty}g_q \Big[ g_{|k-q|}
	\big\{ \sinc[  (k/2-q)\alpha/\pi   ] - \sinc(q\alpha/\pi  ) \big\} 
	\notag
	\\
	&+ g_{k+q} \big\{ \sinc[ (k/2+q)\alpha/\pi   ] - \sinc(q\alpha/\pi)   \big\} \Big]  \bigg\}.
	\label{eq:timeevolutionfourier2}
\end{align}
Due to normalization we have at all times
\begin{align}
	g_0=\frac{1}{2 \pi}.
	\label{eq:zerothcomponent}
\end{align}
We immediately see that $g_{k}=0$ for $k=1, 2, \dots$ is invariant under the time evolution \eqref{eq:timeevolutionfourier2}.
This solution corresponds to a constant angular distribution, that means a completely disordered state.
The stability of the disordered state can be analyzed investigating just the first three modes.
This was done in \cite{RLI14}.
To make this paper self-contained we repeat the stability analysis here.

Taking into account only the first three modes we find the fixed point according to Eq.~\eqref{eq:timeevolutionfourier2} given by the system of equations
\begin{align}
	g_1=& \frac{2}{\eta}\sin\Big( \frac{\eta}{2}\Big)\frac{1}{1+M}(g_1+M g_0 g_1 c_{101} + M g_1 g_2 c_{112} 
	+ M g_2 g_3 c_{123}),
	\label{eq:fixedpointsmodes1}
	\\
	g_2=& \frac{1}{\eta} \sin(\eta)\frac{1}{1+M}(g_2 + M g_0 g_2 c_{202} + M g_{1}^2 c_{211}
	+ M g_{1} g_3 c_{213}),
	\label{eq:fixedpointsmodes2}
	\\
	g_3=& \frac{2}{3\eta} \sin \Big(\frac{3\eta}{2}  \Big) (g_3 + M g_0 g_3 c_{303} + M g_1 g_2 c_{312}),
	\label{eq:fixedpointsmodes3}
\end{align}
where the coupling coefficients $c_{ijk}$ depend on $\alpha$ and are given in Appendix \ref{app:couplingcoefficients}.
From Eqs.~(\ref{eq:fixedpointsmodes1}-\ref{eq:fixedpointsmodes3}) we can derive an equation for the fixed points $\Psi^{*}$ of the order parameter given by Eq.~\eqref{eq:orderparamvsfirstmode}.
Assuming that $\Psi^{*}$ is small and neglecting terms of higher order than $\Psi^{*5}$ we obtain
\begin{align}
	\frac{d_1}{\pi} \Psi^{*} =& M^2\frac{c_{112}c_{211}}{\pi^3 d_2} \Psi^{*3} 
	+ M^4 c_{312}c_{123} c_{211} \frac{c_{211}+c_{112}}{\pi^5 d_3 d_2^2} \Psi^{*5},
	\label{eq:psismallfourier}
\end{align}
where
\begin{align}
	d_1=& \frac{1+M}{\sin(\eta/2)}\frac{\eta}{2}-1-M g_0 c_{101},
	\label{eq:d1}
	\\
	d_2=& \frac{1+M}{\sin(\eta)}\eta-1-M g_0 c_{202},
	\label{eq:d2}
	\\
	d_3=& \frac{1+M}{\sin(3\eta/2)}\frac{3\eta}{2}-1-M g_0 c_{303}.
	\label{eq:d3}
\end{align}
The principal behavior of the system was discussed in \cite{RLI14}.
If $d_1>0$ the fixed point $g_1=0$ is stable and for $d_1<0$ it is unstable.
The critical noise strength $\eta_c$ is determined by the condition $d_1=0$.

If the coefficient of the cubic term in Eq.~\eqref{eq:psismallfourier} is positive the transition at $\eta_c$ appears discontinuous, if it is negative the transition appears continuous
and if it is zero the quintic term has to be considered and the transition is still continuous, where we assumed a homogeneous system in all cases.
Independently on the noise strength, the cubic term vanishes only at one critical point $\alpha_c$. For $\alpha<\alpha_c$ the transition is discontinuous and for $\alpha>\alpha_c$ it is continuous.

For $\alpha>\alpha_c$ we find the leading behavior of the first mode close to $\eta_c$ according to Eq.~\eqref{eq:psismallfourier}
\begin{align}
	\Psi^{*} \approx  \pm D \sqrt{\eta_c-\eta}
	\label{eq:orderparamfouriersmall1}
\end{align}
with
\begin{align}
	D = \frac{\pi}{M} \sqrt{\frac{(1+M) [\sin(\eta_c/2)-\frac{\eta_{c}}{2} \cos(\eta_{c}/2)] d_2(\eta_c)}{-2 c_{112}c_{211}\sin^2(\eta_{c}/2)}}.
	\label{eq:coeffgoneleading1}
\end{align}
At $\alpha=\alpha_c$ the cubic term in Eq.~\eqref{eq:psismallfourier} is zero and the quintic term has to be taken into account, resulting in the leading behavior
\begin{align}
	\Psi^{*} \approx  \pm D' (\eta_c-\eta)^{1/4},
	\label{eq:orderparamfouriersmall2}
\end{align}
where
\begin{align}
	&D' = \frac{\pi}{M} 
	\times \bigg\{ \frac{(1+M)d_2^{2}(\eta_c)d_3(\eta_c) [ \sin(\eta_c/2)-\frac{\eta_c}{2}\cos(\eta_c/2)]}{-2c_{123}c_{211}c_{312}(c_{112}+c_{211})\sin^{2}(\eta_c/2)}  \bigg\}^{1/4}.
	\label{eq:coeffgoneleading2}
\end{align}
Eq.~\eqref{eq:psismallfourier} was given in \cite{RLI14} in different notation, but the coefficients $D$ and $D'$ have not been given explicitly there.
\section{Series ansatz\label{sec:seriesansatz}}
For practical computations we can take into account only finitely many modes. 
This problem can be overcome by taking only a few modes into account exactly and incorporating higher modes by a series ansatz.
\subsection{Geometric series\label{subsec:geo}}
In \cite{RLI14} the following ansatz was proposed. We take the first $l$ modes into account exactly and assume that all higher modes behave as a geometric sequence with parameter
\begin{align}
	\mu=g_{l}/g_{l-1}.
	\label{eq:parametermu}
\end{align}
That means we assume that
\begin{align}
	g_{s}=g_{l-1}\mu^{s-l+1}
	\label{eq:geometricsequence}
\end{align}
for $s\ge l-1$. 

This ansatz is exact in two limiting cases. 
At the critical point $\eta_c$, where all modes starting from $g_1$ are zero, $\mu=0$. For $\eta=0$ all modes are $g_i=\frac{1}{2\pi}$ and thus $\mu=1$.

The big advantage of this ansatz is that the set of Equations \eqref{eq:timeevolutionfourier2} can be closed analytically. 
One obtains after some lengthy calculations, cf. Appendix \ref{app:B}, for even $k$
\begin{align}
	&g_{k}(t+\tau) = \frac{\lambda_k}{1+M} \bigg\{ g_k \Big[ \frac{1}{2} + g_{0} M \big\{ \pi-\alpha +\frac{3}{k}\sin(k\alpha/2) 
	- \frac{1}{2k}\sin(k\alpha)  \big\}     \Big]
	\notag
	\\
	&+\frac{M}{2}\alpha \sum_{q=1}^{l-2+k}\Big[g_q  g_{|k-q|}
	\big\{ \sinc[  (k/2-q)\alpha/\pi   ] - \sinc(q\alpha/\pi  ) \big\} \Big]
	\notag
	\\
	&+ \frac{M}{2}\alpha \sum_{q=1}^{l-2}\Big[g_q g_{k+q} \big\{ \sinc[ (k/2+q)\alpha/\pi   ]
	- \sinc(q\alpha/\pi)   \big\} \Big]
	\notag
	\\
	&+M g_{l-1}^{2}(\mu^2)^{1-l}\Big[ S_{l-1+k/2}(\mu^2, \alpha)
	-\frac{\mu^k}{2}S_{l-1}(\mu^2, \alpha)-\frac{\mu^{-k}}{2}S_{l-1+k}(\mu^2, \alpha)  \Big]
	\bigg\}
	\label{eq:timeevolutionfourier4}
\end{align}
and for odd $k$
\begin{align}
	&g_{k}(t+\tau) = \frac{\lambda_k}{1+M} \bigg\{ g_k \Big[ \frac{1}{2} + g_{0} M \big\{ \pi-\alpha +\frac{3}{k}\sin(k\alpha/2) 
	- \frac{1}{2k}\sin(k\alpha)  \big\}     \Big] 
	\label{eq:timeevolutionfourier5}
	\\
	&+\frac{M}{2}\alpha \sum_{q=1}^{l-2+k}\Big[g_q  g_{|k-q|}
	\big\{ \sinc[  (k/2-q)\alpha/\pi   ] - \sinc(q\alpha/\pi  ) \big\} \Big]
	\notag
	\\
	&+ \frac{M}{2}\alpha \sum_{q=1}^{l-2}\Big[g_q g_{k+q} \big\{ \sinc[ (k/2+q)\alpha/\pi   ]
	- \sinc(q\alpha/\pi)   \big\} \Big]
	\notag
	\\
	&+M g_{l-1}^{2}(\mu^2)^{1-l}\Big[ 2 S_{2l-2+k}(\mu, \frac{\alpha}{2}) - S_{l+\frac{k-1}{2}}(\mu^2, \alpha)
	-\frac{\mu^{k}}{2} S_{l-1}(\mu^2, \alpha)- \frac{\mu^{-k}}{2} S_{l-1+k}(\mu^2, \alpha)  \Big]
	\bigg\},
	\notag
\end{align}
where the functions $S_{i}$ are defined in Appendix~\ref{app:B} and calculated in Appendix~\ref{app:C}.
Thus we obtained a closed time evolution equation for the first $l$ modes. In \cite{RLI14} the series was truncated after $500$ modes and evaluated numerically. With Eqs.~(\ref{eq:timeevolutionfourier4}) and (\ref{eq:timeevolutionfourier5}) we have an analytical treatment of all infinitely many Fourier modes.

\begin{figure}
	\begin{center}
	\includegraphics{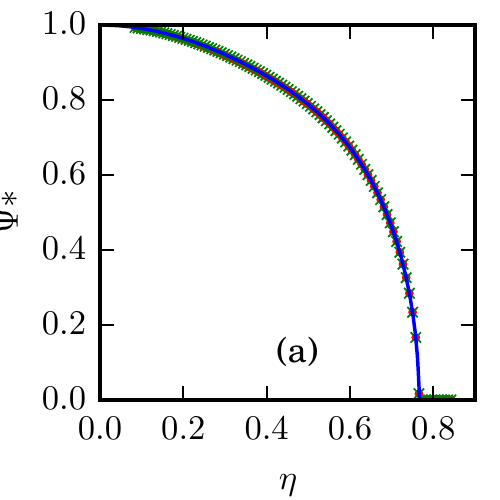}
	\includegraphics{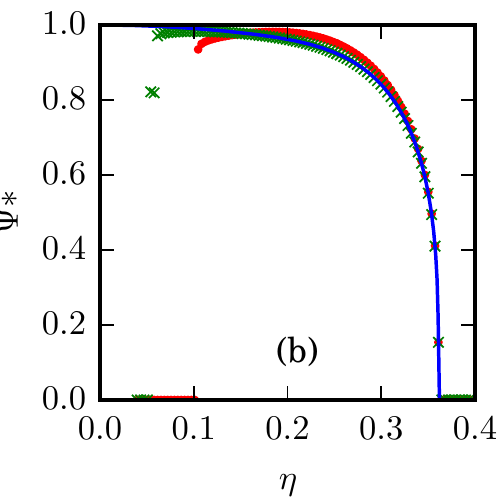}
	\includegraphics{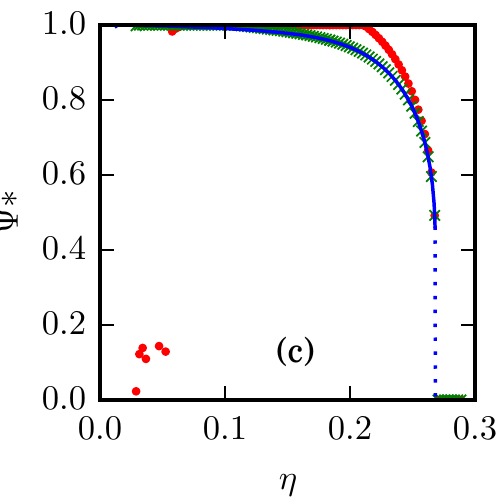}
	\caption[width=0.45\textwidth]{Full Fourier analysis vs. geometric series ansatz. 
	Nonnegative stable fixed points of the order parameter $\Psi$ as a function of noise strength for (a) $\alpha=\pi$, (b) $\alpha=\alpha_c\approx 0.4429096\pi$ and (c) $\alpha=0.35\pi$.
	The blue line indicates the fixed points obtained from the time evolution of $200$ Fourier modes according to Eq.~\eqref{eq:timeevolutionfourier2}, the blue dashed line indicates the jump.
	The symbols show data obtained by taking the first three (red filled circles) or eight (green crosses) modes into account exactly and using a series ansatz for higher modes. For three exact modes and $\alpha=0.35\pi$ (c) there are irregular data for very small noise strengths due to a numerical instability discussed in the last paragraph of subsection \ref{subsec:geo}. 
	Fixed points have been obtained by iterating $10^{5}$ time steps for each point. In all cases $M=0.1$.\label{fig:geo}}
	\end{center}
\end{figure}

In Fig.~\ref{fig:geo} we compare stable fixed points of the order parameter obtained from the full Fourier analysis with those obtained using the geometric series ansatz.
For $\alpha=\pi$ there is very good agreement already for $l=3$.
However, for $\alpha=\alpha_c$ and $\alpha=0.35\pi< \alpha_c$ the geometric approximation breaks down for very small noise.
In Fig.~\ref{fig:geo}c we see irregular data for $l=3$ and small noise, as a result of a numerical instability. The reason is that unphysical distributions are produced. It must always hold $-1\le \langle \cos(k\theta)\rangle \le 1$ and hence $-\frac{1}{\pi} \le g_k \le \frac{1}{\pi}$. However, this condition is violated and we tried to correct this by artificially renormalizing all modes in each step. Apparently this produces numerical instabilities for $l$.
It is not surprising that three modes are not enough to describe the system correct over the full parameter range. They are just essential to cover the transition properly.
By explicitly incorporating the first 8 modes the results are improved but the ansatz is still inaccurate for very small noise.
Therefore in the next subsection we investigate the mode structure for small noise to improve the ansatz.

\subsection{Gaussian decay}

\begin{figure}
	\begin{center}
	\includegraphics{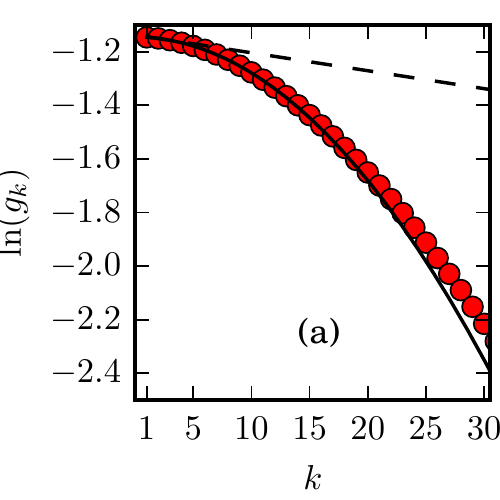}
	\includegraphics{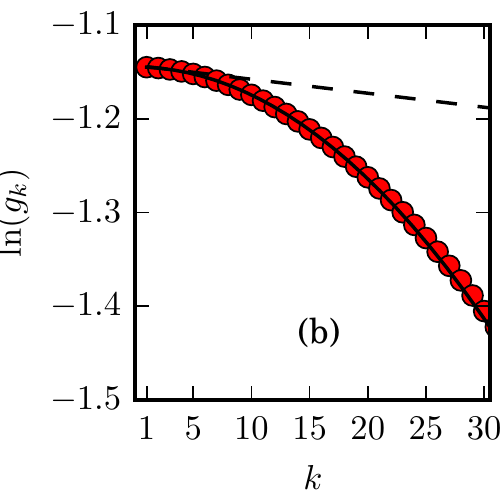}
	\includegraphics{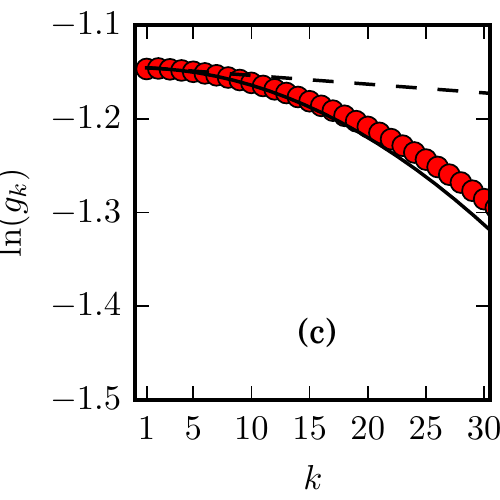}
	\caption{Logarithm of the first $30$ Fourier modes at $\eta=0.05\eta_c$ for (a) $\alpha=\pi$, (b) $\alpha=\alpha_c$ and (c) $\alpha=0.35 \pi$. The red circles are data from the full Fourier analysis of the first $200$ modes, the dashed line is just the straight line through $g_2$ and $g_3$ and the solid line is the Gaussian ansatz \eqref{eq:modeansatz} with parameters obtained from $g_2$ and $g_3$.
	\label{fig:modes}}
	\end{center}
\end{figure}

In Fig.~\ref{fig:modes} we plotted the logarithm of the $k$-th mode vs. $k$ for $\eta= 0.05 \eta_c$.
We find that there is no linear relation but instead the first $10-30$ modes are of the form
\begin{align}
	g_{k} = \exp(a-\gamma k^2).
	\label{eq:modeansatz}
\end{align}
The black solid line in Fig.~\ref{fig:modes} is given by Eq.~\eqref{eq:modeansatz}, where $a$ and $\gamma$ are determined from $g_2$ and $g_3$.
Depending on $\alpha$ it describes the data well for the $10$ to $30$ modes.

The Gaussian decay \eqref{eq:modeansatz} can be understood as follows.
For small noise intensities we expect a narrow distribution of the angular variable.
If the angular distribution is approximated by a sharp Gaussian peak centered around zero the Fourier modes can be asymptotically calculated. One finds that in leading order
\begin{align}
	\frac{g_{k+1}}{g_{k}} -1 \sim  (2k+1) \varepsilon.
	\label{eq:modefraction1}
\end{align}
We obtain the same result from the ansatz \eqref{eq:modeansatz} if $\gamma\sim \varepsilon$ is small.

It is important that the ansatz is a good approximation only where higher modes give an essential contribution to the evolution of the first few modes.
In particular this is the case close to $\eta=0$ where all modes become $1/\pi$.

As in the previous subsection we take into account the first $l$ modes exactly and determine in each step the parameters $a$ and $\gamma$ from Eq.~\eqref{eq:modeansatz} for $k=l$ and $k=l-1$.
Then higher modes are determined according to Eq.~\eqref{eq:modeansatz}.
Unfortunately we have not been able to close the time evolution equation analytically as for the geometric series ansatz.
Therefore we have to truncate the sequence of Fourier modes after the $n$-th mode.
The computational complexity of updating these $n$ modes is reduced to order $n$ instead of order $n^2$ for the full Fourier analysis.

\begin{figure}
	\begin{center}
	\includegraphics{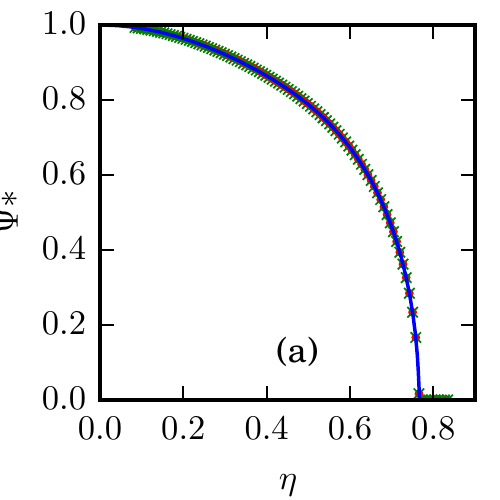}
	\includegraphics{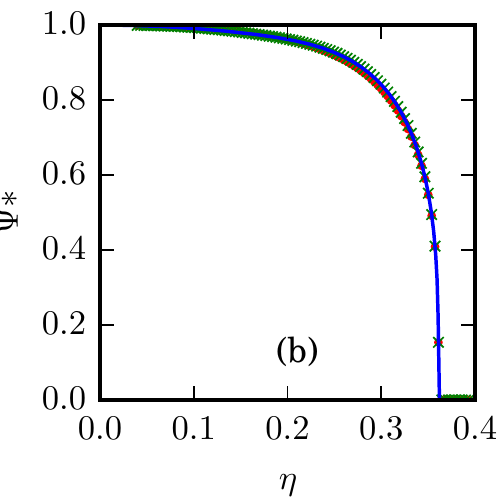}
	\includegraphics{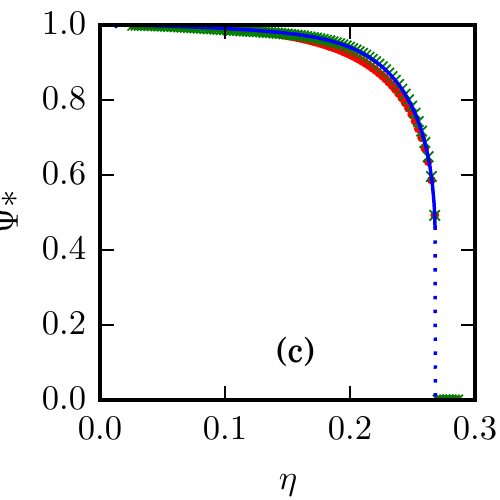}
	\caption{Full Fourier analysis vs. Gaussian decay. 
	Nonnegative stable fixed points of the order parameter $\Psi$ as a function of noise strength for (a) $\alpha=\pi$, (b) $\alpha=\alpha_c\approx 0.4429096\pi$ and (c) $\alpha=0.35\pi$.
	The solid and dotted blue lines are as in Fig.~\ref{fig:geo}.
	The symbols show data obtained by taking the first three (red filled circles) or eight (green crosses) modes into account exactly and using a Gaussian decay ansatz for higher modes. 
	Fixed points have been obtained by iterating $10^{5}$ time steps for each point. In all cases $M=0.1$.\label{fig:series}}
	\end{center}
\end{figure}

In Fig.~\ref{fig:series} we compare the fixed points obtained using the series ansatz with those found by the full Fourier analysis.
Already for $l=3$ the series ansatz gives relatively good results, for $l=8$ there is very good agreement with the full Fourier analysis. In both cases the Fourier series was truncated after the first $200$ modes.
\section{von Mises distribution ansatz\label{sec:vonmises}}

In the recent paper \cite{LSD15} the von Mises distribution was proposed as an ansatz for polar ordering. 
We test this approach in the present system and therefore assume in the following that the distribution function $f(\theta, t)$ is a von Mises distribution at all times given by the single parameter $\Psi(t)$.
The von Mises distribution is given by
\begin{align}
	p_{\Psi}(\theta)= \frac{1}{2\pi I_0[\kappa(\Psi)]}\exp[\kappa(\Psi)\cos(\theta)],
	\label{eq:vonmisesansatz}
\end{align}
where $I_0$ is the modified Bessel-function of the first kind and $\kappa(\Psi)$ is chosen in such a way that
\begin{align}
	\Psi = \int_{0}^{2\pi} \diff \theta \cos(\theta) p_{\Psi}(\theta)= \frac{I_{1}(\kappa(\Psi))}{I_{0}(\kappa(\Psi))}.
	\label{eq:definitionkappa}
\end{align}
Since the right hand side is a strictly monotone function of $\kappa$ that ranges from $-1$ to $+1$ the function $\kappa(\Psi)$ is invertible and well-defined by Eq.~\eqref{eq:definitionkappa}.

If the distribution is chosen according to the von Mises ansatz \eqref{eq:vonmisesansatz} and it evolves according to the time evolution Eq.~\eqref{eq:lowdensitytimeevolution4} it will not be exactly of von Mises form at later times.
However, the hope is that the deviations are not too large and we can assume the following time evolution equation for $\Psi(t)$
\begin{align}
	\Psi(t+\tau) = \int_{0}^{2\pi} \diff \theta \cos(\theta) T[p_{\Psi(t)}(\theta)]=: F[\Psi(t)],
	\label{eq:psitimeevolution}
\end{align}
where $T[\cdot]$ denotes the time evolution of $p(\theta)$ according to Eq.~\eqref{eq:lowdensitytimeevolution4}.

\begin{figure}
	\begin{center}
	\includegraphics{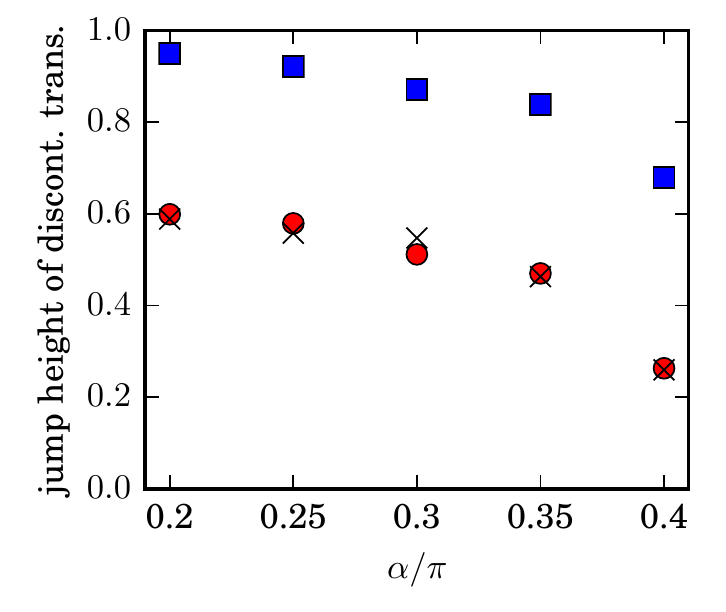}
	\caption{Jump height of the discontinuous transition for $\alpha<\alpha_c$. Data are obtained from the von Mises ansatz (blue squares), the extended von Mises ansatz (red circles) and by considering the first $200$ Fourier modes exactly (black crosses).
	The jump height predicted by the von Mises ansatz deviates by more than $59\%$ from the full Fourier analysis.
	For the extended von Mises ansatz the deviations are at most $7\%$.
	The accuracy of the determined jump height depends highly on the resolution of $\eta$ for data like in Fig.~\ref{fig:geo}c. We chose a resolution of $0.001\eta_c$. Probably the accuracy could be improved using a finer resolution in combination with the iteration of more time steps.
	\label{fig:jumpheight}}
	\end{center}
\end{figure}

\begin{figure}
	\begin{center}
	\includegraphics{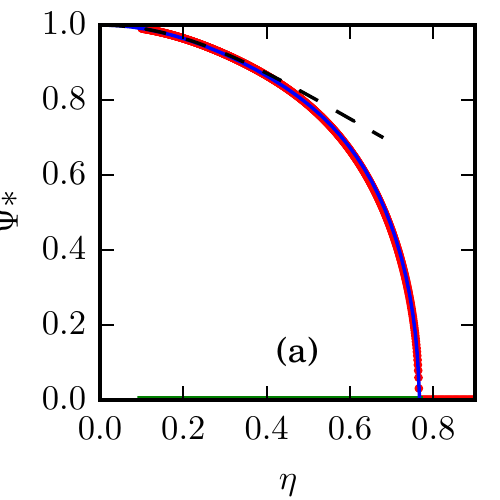}
	\includegraphics{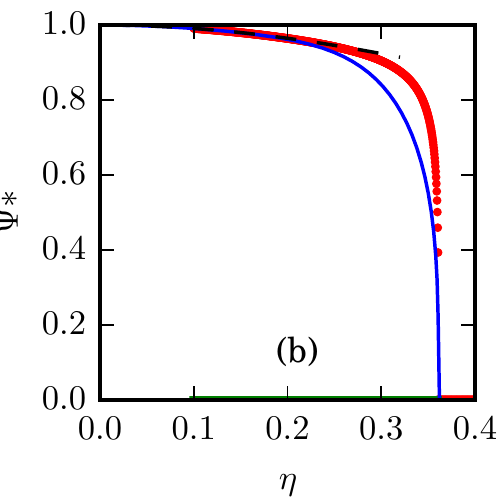}
	\includegraphics{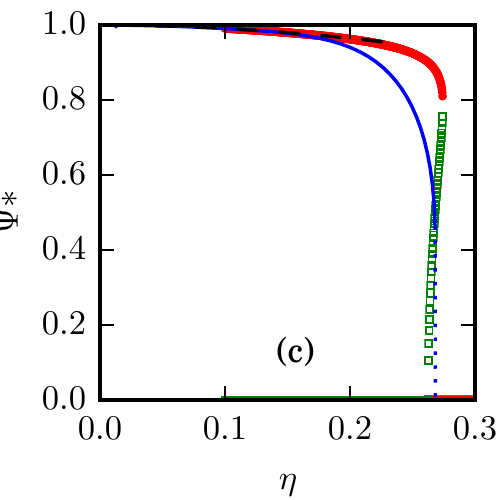}
	\caption{Fourier analysis vs. von Mises ansatz. Nonnegative fixed points of the order parameter $\Psi$ as a function of noise strength for (a) $\alpha=\pi$, (b) $\alpha=\alpha_c\approx 0.4429096\pi$ and (c) $\alpha=0.35\pi$. 
	The solid and dotted blue lines are as in Fig.~\ref{fig:geo}.
	The red filled circles and green empty squares show the numerically obtained stable and unstable fixed points obtained from the von-Mises-distribution ansatz \eqref{eq:vonmisesansatz}, respectively. The dashed black line shows the stable fixed point of obtained from the von Mises ansatz analytically for large order parameters, cf. Eq.~\eqref{eq:fixedpointordered}.
	The stable fixed points of the Fourier mode analysis have been obtained by iterating $10^{5}$ time steps for each point. Stable and unstable fixed points of the von Mises distribution have been obtained using root finding algorithms for the time evolution map \eqref{eq:timeevolutionorderparameter}. In all cases $M=0.1$.\label{fig:vm}}
	\end{center}
\end{figure}

Inserting the ansatz Eq.~\eqref{eq:vonmisesansatz} into Eq.~\eqref{eq:expectationcosine} for $k=1$ we obtain
\begin{align}
	&F(\Psi)= \frac{2}{\eta} \sin\big(\frac{\eta}{2}\big) \frac{1}{1+M} \Bigg\{  \Psi + \frac{M}{4 \pi^2 I_{0}^{2}[\kappa(\Psi)]} 
	\notag
	\\
	&\times
	\bigg[   \int_{0}^{2\pi}\diff \theta \int_{-\alpha}^{\alpha} \diff \hat{\theta} 
	\cos(\theta)
	\exp \Big\{ \kappa(\Psi)\big[ \cos(\theta- \hat{\theta}/2) + \cos(\theta + \hat{\theta}/2)  \big] \Big\}  
	\notag
	\\
	&+  \int_{0}^{2\pi}\diff \theta \int_{[-\pi, -\alpha]\cup [\alpha, \pi]}^{} \diff \hat{\theta} \cos(\theta) 
	\exp \Big\{ \kappa(\Psi)\big[ \cos(\theta) + \cos(\theta + \hat{\theta})  \big] \Big\}  \bigg]       \Bigg\}.
	\label{eq:timeevolutionF1}
\end{align}
The integrals over $\theta$ can be performed
\begin{align}
	&\int_{0}^{2\pi}\diff \theta  
	\cos(\theta)
	\exp \Big\{ \kappa(\Psi)\big[ \cos(\theta- \hat{\theta}/2) + \cos(\theta + \hat{\theta}/2)  \big] \Big\} 
	\notag
	\\
	&= \int_{0}^{2\pi}\diff \theta  
	\cos(\theta)
	\exp \big[ 2\kappa(\Psi) \cos(\theta) \cos(\hat{\theta}/2)  \big] 
	=  2 \pi I_1[2\kappa(\Psi)\cos(\hat{\theta}/2)],
	\label{eq:int1}
\end{align}
\begin{align}
	&\int_{0}^{2\pi}\diff \theta  
	\cos(\theta)
	\exp \Big\{ \kappa(\Psi)\big[ \cos(\theta) + \cos(\theta + \hat{\theta})  \big] \Big\} 
	\notag
	\\
	&= \cos(\hat{\theta}/2) \int_{0}^{2\pi}\diff \tilde{\theta}
	\cos(\tilde{\theta})
	 \exp \Big\{ \kappa(\Psi)\big[ \cos(\tilde{\theta}- \hat{\theta}/2) + \cos(\tilde{\theta} + \hat{\theta}/2)  \big] \Big\} 
	\notag
	\\
	&=  2 \pi  \cos(\hat{\theta}/2) I_1[2\kappa(\Psi)\cos(\hat{\theta}/2)],
	\label{eq:int2}
\end{align}
where we used trigonometric identities and, in the second integral, the substitution $\tilde{\theta} = \theta + \hat{\theta}/2$.
Inserting the integrals \eqref{eq:int1} and \eqref{eq:int2} into Eq.~\eqref{eq:timeevolutionF1} we obtain
\begin{align}
	F(\Psi)
	=&\frac{1}{1+M} \frac{2}{\eta}\sin(\eta/2) 
	\label{eq:timeevolutionorderparameter}
	\\
	&\times \bigg\{ \Psi+ \frac{M}{\pi} \Big[ \int_{0}^{\alpha} \diff \hat{\theta} \frac{I_{1}[2\kappa(\Psi) \cos(\hat{\theta}/2)]}{I_{0}[\kappa(\Psi)]^{2}} 
	+ \int_{\alpha}^{\pi} \diff \hat{\theta} \cos(\hat{\theta}/2) \frac{I_{1}[2\kappa(\Psi) \cos(\hat{\theta}/2)]}{I_{0}[\kappa(\Psi)]^{2}} \Big]      \bigg\}.
	\notag
\end{align}
In Fig.~\ref{fig:vm} we compare the fixed points of the time evolution map \eqref{eq:timeevolutionorderparameter} to those of the Fourier analysis of Sec.~\ref{sec:fourier}.
For $\alpha=\pi$ we find relatively good agreement, however, we discuss in the following subsection that even there, there are quantitative differences even close to the critical point, cf. Fig.~\ref{fig:deviations}, Tab.~\ref{tab:tricrit}.
At the tricritical point there is qualitative agreement with already relatively large quantitative differences and for $\alpha<\alpha_c$ there are even qualitative differences as the position of the jump in the order parameter is not predicted correctly with the von Mises ansatz and there are serious deviations in the jump height of at least $59\%$, cf. Fig.~\ref{fig:jumpheight}.
\subsection{Small order parameter}
We observe that $\Psi=0$ is always a fixed point of \eqref{eq:timeevolutionorderparameter}. 
It describes the state of total disorder and we are investigating the time evolution map \eqref{eq:timeevolutionorderparameter} close to this state.
Differentiating Eq.~\eqref{eq:definitionkappa} several times with respect to $\Psi$, at $\Psi=0$, we obtain the leading behavior of $\kappa(\Psi)$ for small $\Psi$
\begin{align}
	\kappa(\Psi) = 2 \Psi + \Psi^{3} + \frac{5}{6} \Psi^{5} + \mathcal{O}(\Psi^{7}).
	\label{eq:kappasmallpsi}
\end{align}
Inserting this expression into Eq.~\eqref{eq:timeevolutionorderparameter} we can develop the time evolution map for small $\Psi$ yielding
\begin{align}
	&F(\Psi)= \frac{1}{1+M} \frac{2}{\eta}\sin(\eta/2)\bigg\{ 1+\frac{M}{2\pi}c_{101} \Psi + \frac{M}{2\pi} c_{112} \Psi^{3} 
	\label{eq:timeevolutionseries}
	\\
	&+ \frac{M}{2\pi}\Big[- \frac{1}{3} \sin(\alpha) + \frac{4}{9} \sin(3\alpha/2)- \frac{1}{4} \sin(2\alpha) 
	+ \frac{2}{15} \sin(5\alpha/2) - \frac{1}{18}\sin(3\alpha) \Big]\Psi^{5} \! + \! \mathcal{O}(\Psi^7) \bigg\},
	\notag
\end{align}
where the coefficients $c_{101}$ and $c_{112}$ are given by Eqs.~\eqref{eq:c101} and \eqref{eq:c112}.
Since $F(\Psi)$ is an odd function there appear no even powers.

The disordered solution $\Psi=0$ changes stability when $F'(0)=1$.
Hence we find the critical noise strength $\eta_c$ according to Eq.~\eqref{eq:timeevolutionseries} by
\begin{align}
	&1 = \frac{2}{\eta_{c}} \sin(\eta_{c}/2) \frac{1}{1+M}
	\bigg\{1 + M \Big[\frac{4}{\pi} \sin(\alpha/2) + 1 - \frac{\alpha}{\pi} -\frac{1}{\pi}\sin(\alpha) \Big] \bigg\}.
	\label{eq:criticalnoise}
\end{align}
The critical noise strength \eqref{eq:criticalnoise} predicted by the von Mises ansatz coincides with the one predicted by the Fourier analysis, cf. \cite{RLI14}.

Since $F'(0)$  decreases monotone as $\eta$ increases we find that
\begin{align}
	F'(0) \gtrless 1 \text{ if } \eta \lessgtr \eta_{c}.
	\label{eq:fprimeetac}
\end{align}
Hence the disordered solution $\Psi=0$ is stable for $\eta > \eta_{c}$ and it is unstable for $\eta<\eta_{c}$.

The transition at $\eta_{c}$ is continuous if $F'''(0)|_{\eta=\eta_{c}}>0$ and discontinuous if $F'''(0)|_{\eta=\eta_{c}}<0$.
The sign of $F'''(0)$ equals the sign of $\sin(3\alpha/2) -\frac{3}{4}\sin(\alpha)-\frac{3}{8}\sin(2\alpha) $.
It is positive if $\alpha<\alpha_c$ and it is negative if $\alpha>\alpha_c$.
At $\alpha=\alpha_c\approx 0.4429096\pi$ $F'''(0)=0$.

Thus the tricritical point is given by $\eta=\eta_{c}$ and $\alpha=\alpha_{c}$.
The von Mises ansatz leads to the same tricritical point as the full Fourier analysis, cf. \cite{RLI14}.
This means that $F'(0)$ and $F'''(0)$ of both descriptions become zero for the same parameters. However, this does not imply that $F'(0)$ and $F'''(0)$ of both descriptions coincide for all parameters. 

\begin{figure}
	\begin{center}
	\includegraphics{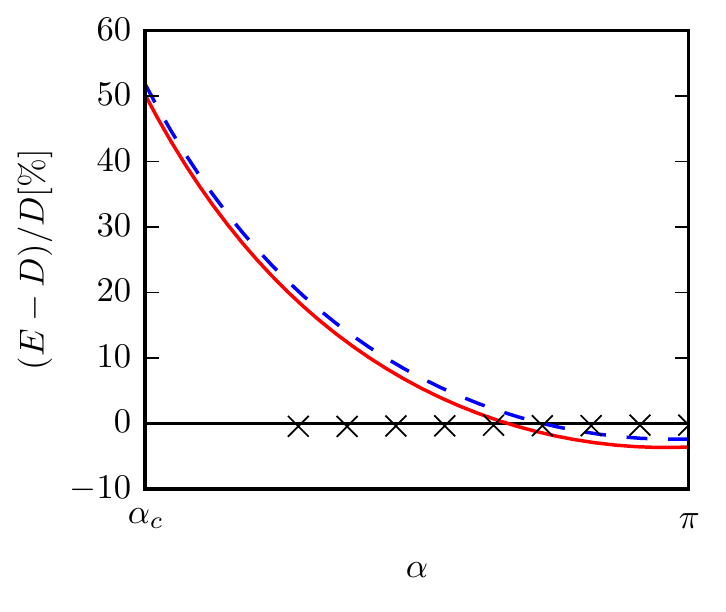}
	\caption{Deviations between the order parameter $\Psi=D(\eta_c-\eta)^{1/2}$ close to $\eta_c$ and the prediction by the von Mises ansatz by $\Psi=E(\eta-\eta_c)^{1/2}$ as a function of $\alpha$ for $M=0.1$ (solid red line) and $M=0.01$ (dashed blue line) and a numerical test of the extended von Mises ansatz (black crosses). For the von Mises ansatz the deviations become huge close to $\alpha\approx \alpha_c$ and also for the standard VM at $\alpha=\pi$ they are larger than $3\%$. The extended von Mises ansatz predicts exactly the same value of $D$ as the full system analysis. With the extended ansatz we determined the order parameter at $\eta= \eta_c - 0.01 \eta_c, \eta_c - 0.02 \eta_c, \eta_c-0.04\eta_c, \eta_c-0.08\eta_c$ numerically and used a single parameter fit to obtain the coefficient $D$ for $M=0.1$. The deviations are about $0.5\%$. The solid black line at zero is just a guide to the eye.\label{fig:deviations}}
	\end{center}
\end{figure}

We calculate the leading behavior of the order parameter close to the critical point for $\alpha>\alpha_c$ and $\alpha=\alpha_c$.
Above the tricritical point it suffices to take into account terms up to $\Psi^3$ in Eq.~\eqref{eq:timeevolutionseries}.
Hence we find the fixed points $\Psi^{*}$ of the time evolution map as solutions of
\begin{align}
	d_1 \Psi^{*} = \frac{M}{2\pi} c_{112} \Psi^{*3},
	\label{eq:orderparamdev1}
\end{align}
where $d_1$ is given by Eq.~\eqref{eq:d1} and $c_{112}$ by Eq.~\eqref{eq:c112}.
The nonzero solutions close to the critical point are
\begin{align}
	\Psi^{*} =& \pm E \sqrt{\eta_c-\eta} 
	\label{eq:leadingorderparamvm1}
\end{align}
with
\begin{align}
	E =\sqrt{\frac{-\pi(1+M) [\sin(\eta_c/2)-\frac{\eta_{c}}{2} \cos(\eta_{c}/2)] }{M c_{112}\sin^2(\eta_{c}/2)}}
	\label{eq:leadingorderparamvmcoeff1}
\end{align}
The coefficient $E$ predicted by the von Mises ansatz differs from the corresponding coefficient $D$ that is obtained by Fourier analysis, cf. Eq.~\eqref{eq:coeffgoneleading1}.
In Fig.~\ref{fig:deviations} we show the deviations as a function of $\alpha$.
Close to $\alpha_c$ they become larger than $50\%$ but also for the standard Vicsek model, at $\alpha=\pi$, there are deviations of more than three percent.
At the tricritical point there is no term proportional to $\Psi^{3}$ in Eq.~\eqref{eq:timeevolutionseries} and the next order term has to be taken into consideration resulting in
\begin{align}
	\Psi^{*} =& \pm E' (\eta_c-\eta)^{1/4}
	\label{eq:leadingorderparamvm2}
\end{align}
with
\begin{align}
	E' =& \bigg\{-\pi\frac{1+M}{M}\frac{\sin(\eta_c/2)-\frac{\eta_c}{2}\cos(\eta_c/2)}{\sin^2(\eta_c/2)}
	\label{eq:leadingorderparamvmcoeff2}
	\\
	&\bigg/\Big[ - \frac{1}{3} \sin(\alpha) + \frac{4}{9} \sin(3\alpha/2) 
	- \frac{1}{4} \sin(2\alpha)+ \frac{2}{15} \sin(5\alpha/2) - \frac{1}{18}\sin(3\alpha) \Big]  \bigg\}^{1/4}.
	\notag
\end{align}
In Table~\ref{tab:tricrit} we compare the coefficient $E'$ predicted by the von Mises ansatz with the coefficient $D'$ obtained by Fourier analysis, cf. Eq.~\eqref{eq:coeffgoneleading2}. There are huge deviations of about $90\%$. Thus the von Mises ansatz is not able to appropriately determine the order parameter close to the critical point.
\begin{table}
	\begin{center}
\begin{tabular}{l|c|c|c}
	$M$ & $D'$ & $E'$ & $\frac{|D'-E'|}{D'}[\%]$  \\
	&&&\vspace{-0.3cm}\\
	\hline
	0.1 & 1.59389234612 & 3.00658504094 & 88.6 \\
	\hline
	0.01 & 2.06282992919 & 3.96175328257 & 92.1
\end{tabular}
\caption{Leading behavior of the order parameter close to the critical point $\eta_c$ for $\alpha=\alpha_c$, $\Psi=D'(\eta_c-\eta)^{1/4}$, cf. Eq.~\eqref{eq:coeffgoneleading2}, and approximation by von Mises ansatz $\Psi=E'(\eta_c-\eta)^{1/4}$, cf. Eq.~\eqref{eq:leadingorderparamvmcoeff2}. The coefficient predicted by the von Mises ansatz deviates about $90\%$ from the correct value.\label{tab:tricrit}}
	\end{center}
\end{table}

\subsection{Large order parameter}
For very small noise strength we are deep in the ordered phase and it becomes difficult to accurately evaluate Eq.~\eqref{eq:timeevolutionorderparameter} numerically.
However, for small noise and close to the stable fixed points we can approximate Eq.~\eqref{eq:timeevolutionorderparameter} analytically.
One of the stable fixed points of the order parameter $\Psi^{*}_{+} \rightarrow 1$ as $\eta \rightarrow 0$, hence in this limit $1-\Psi \ll1$. We can develop the time evolution map \eqref{eq:timeevolutionseries} for $1-\Psi \ll 1$ and obtain
\begin{align}
	F(\Psi) \approx \frac{2}{\eta} \sin\Big(\frac{\eta}{2}\Big) \frac{1}{1\!+\!M}\Big[\Psi\! +\!\frac{M}{2}(1\!+\!\Psi)\Big]\! +\! \mathcal{O}[(1-\Psi)^{2}].
	\label{eq:timeevolutionmapordered}
\end{align}
This leads to the fixed point
\begin{align}
	\Psi^{*}_{+}= \frac{1}{\eta}\sin\Big(\frac{\eta}{2}\Big)\frac{M}{1+M}\bigg/ \bigg[1- \frac{2}{\eta} \sin(\eta/2) \frac{1+M/2}{1+M} \bigg].
	\label{eq:fixedpointordered}
\end{align}
Due to symmetry there is another stable fixed point at $\Psi^{*}_{-}=-\Psi^{*}_{+}$.

\section{Extended von Mises distribution ansatz\label{sec:extendedvonmises}}

To further improve upon the results of the previous section we extend the von Mises-distribution ansatz by adding another term
\begin{align}
	p(\theta, t) =& \frac{1}{Z}\bigg\{ \exp\Big[A \cos(\theta)\Big] + B \exp\Big[C \cos(2\theta) \Big] \bigg\},
	\label{eq:extendedvonmises}
	\\
	Z=& \int_{0}^{2\pi} \diff \theta \bigg\{ \exp\Big[A \cos(\theta)\Big] + B \exp\Big[C \cos(2\theta) \Big] \bigg\}
	= 2\pi\bigg[ I_{0}(A) + B I_{0}(C) \bigg].
	\label{eq:normextendedvonmises}
\end{align}
The first term is just the von Mises distribution. The second term allows corrections that take into account the next two modes.
As in Sec.~\ref{sec:vonmises} we assume that the distribution function has the functional form \eqref{eq:extendedvonmises} at all times.
Hence the parameters $A$, $B$, $C$ and the normalization factor $Z$ depend on time.
The time evolution of these quantities is obtained in two steps.

In the first step we calculate $\langle \cos(\theta)\rangle$, $\langle \cos(2\theta)\rangle$ and $\langle \cos(3\theta)\rangle$ at time $t+\tau$ according to Eq.~\eqref{eq:expectationcosine} assuming that $f(\theta, t)$ is given by Eq.~\eqref{eq:extendedvonmises}.
The integrals in Eq.~\eqref{eq:expectationcosine} have to be solved numerically. Thus two-dimensional numerical integration is required in the first step.

In the second step on finds the time evolved parameters $A(t+\tau)$, $B(t+\tau)$ and $C(t+\tau)$ that are compatible with $\langle \cos(\theta)\rangle(t+\tau)$, $\langle \cos(2\theta)\rangle(t+\tau)$ and $\langle \cos(3\theta)\rangle(t+\tau)$ according to the ansatz Eq.~\eqref{eq:extendedvonmises} which leads to
\begin{align}
	\Psi = \langle \cos(\theta) \rangle =& \frac{2\pi}{Z} I_{1}(A),
	\label{eq:firstmodeextendedvonmises}
	\\
	\langle \cos(2\theta) \rangle =& \frac{2\pi}{Z} \Big[ I_{2}(A)+ B I_{1}(C) \Big],
	\label{eq:secondmodeextendedvonmises}
	\\
	\langle \cos(3\theta) \rangle =& \frac{2\pi}{Z} I_{3}(A).
	\label{eq:thirdmodeextendedvonmises}
\end{align}
The parameter $A$ can be obtained easily as from Eqs.\eqref{eq:firstmodeextendedvonmises} and \eqref{eq:thirdmodeextendedvonmises} follows
\begin{align}
	\frac{\langle \cos(3\theta)\rangle}{\langle \cos(\theta)\rangle} =  \frac{I_{3}(A)}{I_{1}(A)}.
	\label{eq:findA}
\end{align}
This equation has a unique solution as the right side is strictly monotone in $A$ and ranges from $-\infty$ to $\infty$.
Thus $A(t+\tau)$ can be calculated numerically with high efficiency.

From Eqs.~\eqref{eq:firstmodeextendedvonmises} and \eqref{eq:thirdmodeextendedvonmises} we find that
\begin{align}
	\frac{I_{1}(A)\langle \cos(2\theta)\rangle-I_{2}(A)\langle \cos(\theta)\rangle}{I_{1}(A)-I_{0}(A)\langle \cos(\theta)\rangle}= \frac{I_{1}(C)}{I_{0}(C)}.
	\label{eq:findC}
\end{align}
We can numerically solve this equation for $C$ as the right hand side is again strictly monotone in $C$.

Having found $A$ and $C$ one obtains $B$ for example from Eqs.~\eqref{eq:normextendedvonmises} and \eqref{eq:firstmodeextendedvonmises} as
\begin{align}
	B= \frac{ I_{1}(A)/\langle \cos(\theta)\rangle -I_{0}(A)  }{  I_{0}(C)  }.
	\label{eq:findB}
\end{align}
The computationally most expensive step in time evolving the distribution \eqref{eq:extendedvonmises} is the two dimensional integral in Eq.\eqref{eq:expectationcosine} but it is still relatively fast. 

\begin{figure}
	\begin{center}
	\includegraphics{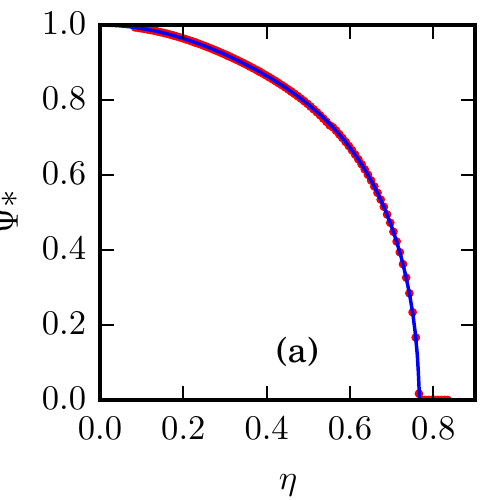}
	\includegraphics{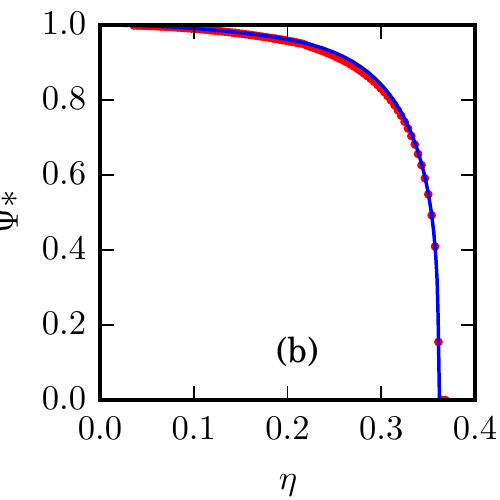}
	\includegraphics{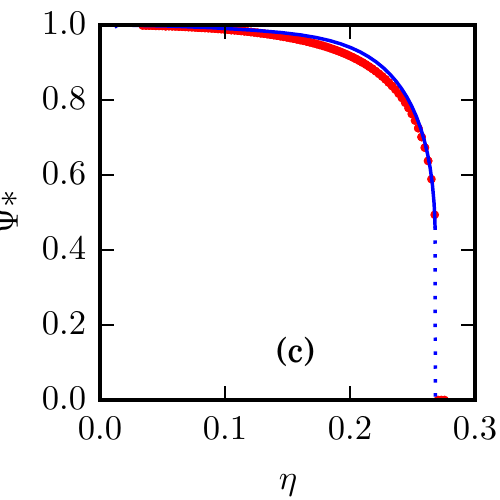}
	\caption{Fourier analysis vs. extended von Mises ansatz. Nonnegative stable fixed points of the order parameter $\Psi$ as a function of noise strength for (a) $\alpha=\pi$, (b) $\alpha=\alpha_c\approx 0.4429096\pi$ and (c) $\alpha=0.35\pi$. 
	The solid and dotted blue lines are as in Fig.~\ref{fig:geo}.
	The red filled circles show the fixed points obtained from the time evolution of the extended von Mises-distribution \eqref{eq:extendedvonmises}. Fixed points have been obtained by iterating $10^{5}$ time steps for each point. In all cases $M=0.1$.\label{fig:extendedvm}}
	\end{center}
\end{figure}

In Fig.~\ref{fig:extendedvm} we compare stable fixed points obtained with the extended von Mises ansatz \eqref{eq:extendedvonmises} with the one resulting from the full Fourier analysis considering the first $200$ modes.
For the standard Vicsek model, $\alpha=\pi$, there is perfect agreement.
In the other two considered cases, $\alpha=\alpha_c$ and $\alpha=0.35\pi<\alpha_c$ there are still slight deviations for intermediate noise strengths, however, the results are much better than for the pure von Mises ansatz, cf. Fig.~\ref{fig:vm}. Also the jump height of the discontinuous transition is predicted much better with deviations of at most $7\%$, cf. Fig.~\ref{fig:jumpheight}.

To analyze the leading behavior of the order parameter close to the critical point we make the ansatz that the first three modes behave as
\begin{align}
	\Psi=\langle \cos (\theta)\rangle =: \varepsilon,
	\qquad
	\langle \cos(2\theta)\rangle \sim \varepsilon^2,
	\qquad
	\langle \cos(3\theta)\rangle \sim \varepsilon^3,
	\label{eq:exvmansatz3}
\end{align}
where $\varepsilon$ is small. This leads according to Eqs.~\eqref{eq:findA} and \eqref{eq:findC} in leading order to
\begin{align}
	A \sim \varepsilon,
	\qquad
	B  \sim \varepsilon^2.
	\label{eq:exvmAB}
\end{align}
One finds that all higher modes are of order $\varepsilon^4$ or of higher order. 
Thus taking into account only terms up to $\varepsilon^3$ just the first three modes need to be considered.
It is possible to develop $A, B$ and $C$ for small $\varepsilon$ and evaluate the time evolution equation \eqref{eq:expectationcosine} for the first three modes taking into account only terms up to $\varepsilon^3$.
One arrives exactly at Eqs.~(\ref{eq:fixedpointsmodes1}-\ref{eq:fixedpointsmodes3}), where
\begin{align}
	\langle \cos(k \theta)\rangle = \pi g_k.
	\label{eq:moderel}
\end{align}
Thus the extended von Mises ansatz is able to predict the leading behavior of the order parameter close to the disordered phase correctly.
In Fig.~\ref{fig:deviations} we compare the predicted leading behavior with numerical results. The deviations are about $0.5\%$ and seem to not depend on parameters. The reason for these deviations can be that the numerical data have been obtained not close enough to the critical point. Also numerical inaccuracies in the numerical integration of Eq.~\eqref{eq:expectationcosine} can play a role.

\section{Conclusions\label{sec:conclusions}}
We investigated polar ordering in the two-dimensional Vicsek model with bounded confidence interactions where each particle interacts only with particles whose direction differs by no more than $\alpha$. For $\alpha=\pi$ the standard Vicsek model is recovered.
We investigated only homogeneous systems.
In this case there is a critical value $\alpha_c$ such that for $\alpha>\alpha_c$, assuming all states are homogeneous, there is a continuous state transition at noise strength $\eta_c$ such that the system is disordered for $\eta>\eta_c$ and there is polar order for $\eta<\eta_c$.
For $\alpha<\alpha_c$ the transition becomes discontinuous.

Assuming molecular chaos and low particle densities we derived an explicit algebraic time evolution equation of the angular distribution in Fourier space, Eq.~\eqref{eq:timeevolutionfourier2}.
In principle this equation was given already in \cite{RLI14}, however several coupling coefficients have not been calculated explicitly, which we did in this paper.
The computational complexity of time evolving the first $n$ Fourier modes is of order $n^2$.
To simplify the problem from infinite dimensions we tested several approaches.

In an ansatz proposed in \cite{RLI14} only the first $l$ modes are considered exactly and all higher modes are assumed to decay like a geometric series.
This approach has the advantage that all infinitely many modes can be considered analytically and we achieve a closed time evolution equation for the first $l$ modes, Eqs.~\eqref{eq:timeevolutionfourier4} and \eqref{eq:timeevolutionfourier5}.
The description is very accurate at the critical point and close to it. It is good when the order parameter is not to large. For the standard Vicsek model the description is good for all noise strengths. However, for small $\alpha$ and small noise strength there are serious deviations from the real behavior.

For small noise the angular distribution is a sharp peak and as a consequence we find that the Fourier modes do not decay like a geometric sequence but like a Gaussian. 
Using a Gaussian decay ansatz for higher Fourier modes there is very good agreement with the real system over the full parameter range.
Unfortunately, in this case we have not been able to close the time evolution equation including all (infinitely many) modes, such that the Fourier series has to be truncated after finitely many modes like in the full analysis. However, the computational complexity to evolve the first $n$ modes is reduced to the order $n$.

The so far mentioned approaches used a series ansatz in Fourier space. We also tried a more direct recently proposed \cite{LSD15} ansatz in real space. 
We assume that the angular distribution is of von Mises type at all times. 
The big advantage is that the von Mises distribution has only a single parameter.
Hence the time evolution of the distribution is reduced to a one-dimensional map that evolves the order parameter.
This ansatz is able to qualitatively reproduce the correct system behavior and also the correct critical point.

However, the von Mises ansatz is too simple to give a quantitatively correct description of the system.
There are serious deviations even for the standard Vicsek model.
The order parameter close to the critical point deviates about $3\%$ for $\alpha=\pi$ and more than $50\%$ close to the tricritical point $\alpha_c$.
The position and the jump height of the discontinuous transition is not predicted properly.
The ansatz is only able to handle the first Fourier mode correctly, however once the first mode is chosen, also all higher modes are determined and there is no more freedom.
The correct handling of the first mode is enough to predict the critical point but to also achieve the correct order parameter at least one more degree of freedom is necessary to control the second mode.
At the tricritical point also the third mode is necessary.

To combine the advantage of a simple angular distribution ansatz with the necessity of the correct handling of at least the first three Fourier modes we extended the von Mises ansatz by adding another term.
In that way the order parameter close to the critical point is predicted correctly.
For the standard Vicsek model there is almost perfect agreement of the order parameter over the full parameter range.
Also for smaller values of $\alpha$ there is relatively good agreement although there are slight deviations.
For not too small $\alpha$, the jump height of the discontinuous transitions is predicted correctly.

\begin{appendix}
\section{Time evolution in Fourier space\label{app:coupling}}
The coupling coefficients are given by
\begin{align}
	B_{kpq}(\alpha)=& \frac{1}{(2\pi)^{2}}\int_{0}^{2\pi}\int_{0}^{2\pi} \diff \theta_1 \diff \theta_2 
	\cos(k \Phi_1)\cos(p\theta_1)\cos(q\theta_2)
	= B_{kpq}^{(1)}+B_{kpq}^{(2)},
	\label{eq:couplingcoeff}
\end{align}
where
\begin{align}
	B_{kpq}^{(1)}=&\frac{1}{(2\pi)^{2}}\int_{0}^{2\pi}\diff \theta_1 \cos(k\theta_1) \cos(p\theta_1) 
	\int_{\theta_1+\alpha}^{\theta_1-\alpha+2\pi} \diff \theta_2 \cos(q\theta_2)
	\notag
	\\
	=& \begin{cases}
		-\frac{1}{4\pi q}\sin(q\alpha)(a_{kpq} + b_{kpq}) \text{ for }q\neq 0\\
		\frac{1}{2}\Big(1-\frac{\alpha}{\pi}\Big)\delta_{0}(k-p) \text{ for }q=0, k\neq 0\\
		1-\frac{\alpha}{\pi} \text{ for } k=p=q=0,
	\end{cases}
	\\
	B_{kpq}^{(2)}=&\frac{1}{(2\pi)^{2}} \int_{0}^{2\pi} \diff \theta_1\cos(p \theta_1) 
	\int_{\theta_1-\alpha}^{\theta_1+\alpha}\diff \theta_2 \cos[k\Phi_1(\theta_1,\theta_2)]\cos(q\theta_2)
	\label{eq:couplingcoeff2}
	\\
	=&\frac{\alpha}{4\pi}\Big[\sinc\Big(\frac{p+q}{2\pi}\alpha\Big)a_{kpq} + \sinc\Big(\frac{p-q}{2\pi}\alpha\Big)b_{kpq}        \Big],
	\notag
\end{align}
where
\begin{align}
	\sinc(x)= 
	\begin{cases}
		\frac{\sin(x\pi)}{x\pi} \text{ for }x\neq 0\\
		1 \text{ for } x=0
	\end{cases}
	\label{eq:sinc}
\end{align}
and
\begin{align}
	a_{kpq}&= \delta_{0}(k+p-q) + \delta_{0}(k-p+q),
	\\
	b_{kpq}&= \delta_{0}(k+p+q) + \delta_{0}(k-p-q),
	\label{eq:deltacoeffs}
\end{align}
where
\begin{align}
	\delta_{0}(x)=
	\begin{cases}
		1 \text{ for } x=0\\
		0 \text{ else}.
	\end{cases}
	\label{eq:deltazero}
\end{align}
Inserting the coupling coefficients \eqref{eq:couplingcoeff2} into the time evolution Eq.~\eqref{eq:timeevolutionfourier} we obtain for $k>0$ the time evolution Eq.~\eqref{eq:timeevolutionfourier2}.

\section{Coupling coefficients \label{app:couplingcoefficients}}
\begin{align}
	c_{101}=& 2\pi -2\alpha + 8 \sin \Big( \frac{\alpha}{2} \Big) - 2 \sin(\alpha),
	\label{eq:c101}
	\\
	c_{112}=& \frac{4}{3} \sin\Big( \frac{3\alpha}{2} \Big) - \frac{1}{2} \sin (2\alpha) - \sin (\alpha),
	\label{eq:c112}
	\\
	c_{123}=& \frac{4}{5} \sin\Big( \frac{5\alpha}{2} \Big) - \frac{1}{3} \sin(3\alpha) - \frac{1}{2} \sin(2\alpha),
	\label{eq:c123}
	\\
	c_{202}=& 2\pi -2\alpha + 4 \sin(\alpha) - \sin(2\alpha),
	\label{eq:c202}
	\\
	c_{211}=& \alpha - \sin(\alpha),
	\label{eq:c211}
	\\
	c_{213}=& \sin(2\alpha) - \frac{1}{3} \sin(3\alpha) - \sin(\alpha),
	\label{eq:c213}
	\\
	c_{303}=&2\pi -2\alpha + \frac{8}{3} \sin \Big( \frac{3\alpha}{2}\Big) - \frac{2}{3} \sin(3\alpha),
	\label{eq:c303}
	\\
	c_{312}=& 4 \sin(\frac{\alpha}{2}) - \sin(\alpha) - \frac{1}{2} \sin(2\alpha).
	\label{eq:c312}
\end{align}
\vfill
\section{Closure of time evolution\label{app:B}}

With the geometric series ansatz for modes higher or equal to $l$ Eq.~\eqref{eq:timeevolutionfourier2} can be written as
\begin{align}
	&g_{k}(t+\tau) = \frac{\lambda_k}{1+M} \bigg\{ g_k \Big[ \frac{1}{2} + g_{0} M \big\{ \pi-\alpha +\frac{3}{k}\sin(k\alpha/2) 
	- \frac{1}{2k}\sin(k\alpha)  \big\}     \Big]
	\notag
	\\
	&+\frac{M}{2}\alpha \sum_{q=1}^{l-2+k}\Big[g_q  g_{|k-q|}
	\big\{ \sinc[  (k/2-q)\alpha/\pi   ] - \sinc(q\alpha/\pi  ) \big\} \Big]
	\notag
	\\
	&+ \frac{M}{2}\alpha \sum_{q=1}^{l-2}\Big[g_q g_{k+q} \big\{ \sinc[ (k/2+q)\alpha/\pi   ] 
	- \sinc(q\alpha/\pi)   \big\} \Big] 
	\notag
	\\
	&+\frac{M}{2}\alpha g_{l-1}^{2}\sum_{q=l-1+k}^{\infty}\Big[ \mu^{2q-2l+2-k}\big\{ \sinc[  (k/2-q)\alpha/\pi   ] 
	-\sinc(q\alpha/\pi  ) \big\} \Big]
	\notag
	\\
	&+ \frac{M}{2}\alpha g_{l-1}^{2} \sum_{q=l-1}^{\infty}\Big[\mu^{2q-2l+2+k} \big\{ \sinc[ (k/2+q)\alpha/\pi   ] 
	-\sinc(q\alpha/\pi)   \big\} \Big]  \bigg\}
	.
	\label{eq:timeevolutionfourier3}
\end{align}
We introduce infinite sums of the type
\begin{align}
	S_{k}(\mu, \alpha):= \sum_{q=k}^{\infty}\mu^{q}\frac{\sin(q\alpha)}{q},
	\label{eq:fmu1}
\end{align}
that can be performed, cf. Appendix \ref{app:C}, resulting in
\begin{align}
	S_{k}(\mu, \alpha):= \frac{1}{2} \arctan(y,x) - \sum_{q=1}^{k-1}\mu^{q}\frac{\sin(q\alpha)}{q},
	\label{eq:fmu2}
\end{align}
with
\begin{align}
	x=&\frac{(1-\mu\cos\alpha)^2 -\mu^2\sin^2\alpha}{ (1-\mu\cos\alpha)^2 +\mu^2\sin^2\alpha},
	\label{eq:xy1}
	\\
	y=&\frac{2(1-\mu\cos\alpha)\mu\sin\alpha}{ (1-\mu\cos\alpha)^2 +\mu^2\sin^2\alpha}.
	\label{eq:xy2}
\end{align}
We find for even $k$
\begin{align}
	\sum_{q=l-1+k}^{\infty}\mu^{(2q-2l+2-k)}\frac{\sin[(q-k/2)\alpha]}{(q-k/2)\alpha}
	&=\frac{1}{\alpha}(\mu^2)^{1-l}\sum_{s=l-1+k/2}^{\infty} (\mu^{2})^{s}\frac{\sin(s\alpha)}{s}
	\notag
	\\
	&=\frac{1}{\alpha}(\mu^2)^{1-l} S_{l-1+k/2}(\mu^{2}, \alpha),
\end{align}
\begin{align}
	&\sum_{q=l-1+k}^{\infty}\mu^{(2q-2l+2-k)}\frac{\sin(q\alpha)}{q\alpha}
	=\frac{1}{\alpha}(\mu^2)^{1-l-k/2} S_{l-1+k}(\mu^{2}, \alpha),
\end{align}
\begin{align}
	\sum_{q=l-1}^{\infty}\mu^{(2q-2l+2+k)}\frac{\sin[(q+k/2)\alpha]}{(q+k/2)\alpha}
	&=\frac{1}{\alpha}(\mu^2)^{1-l}\sum_{s=l-1+k/2}^{\infty} (\mu^{2})^{s}\frac{\sin(s\alpha)}{s}
	\notag
	\\
	&=\frac{1}{\alpha}(\mu^2)^{1-l} S_{l-1+k/2}(\mu^{2}, \alpha),
\end{align}
\begin{align}
	&\sum_{q=l-1}^{\infty}\mu^{(2q-2l+2+k)}\frac{\sin(q\alpha)}{q\alpha}
	=\frac{1}{\alpha}(\mu^2)^{1-l+k/2} S_{l-1}(\mu^{2}, \alpha).
	\label{eq:sumsevenk}
\end{align}
Inserting these expressions into Eq.~\eqref{eq:timeevolutionfourier3} we obtain Eq.~\eqref{eq:timeevolutionfourier4}.

On the other hand for odd $k$ we have
\begin{align}
	&\sum_{q=l-1+k}^{\infty} \mu^{2q-2l+2-k}\frac{\sin[(q-k/2)\alpha]}{(q-k/2)\alpha}
	=\frac{2}{\alpha}(\mu^{2})^{1-l}\sum_{q=l-1+k}^{\infty}\mu^{2q-k} \frac{\sin[(2q-k)\frac{\alpha}{2}]}{2q-k}
	\notag
	\\
	&=\frac{2}{\alpha}(\mu^{2})^{1-l}\sum_{s=2l-2+k, s \text{ odd}}^{\infty}\mu^{s} \frac{\sin(s\frac{\alpha}{2})}{s}
	\notag
	\\
	&=\frac{2}{\alpha}(\mu^{2})^{1-l}\Big[ \sum_{s=2l-2+k}^{\infty}\mu^{s} \frac{\sin(s\frac{\alpha}{2})}{s} 
	- \sum_{s=2l-1+k, s \text{ even}}^{\infty}\mu^{s} \frac{\sin(s\frac{\alpha}{2})}{s} \Big]
	\notag
	\\
	&=\frac{2}{\alpha}(\mu^{2})^{1-l}\Big[ S_{2l-2+k}(\mu, \frac{\alpha}{2}) - \frac{1}{2}\sum_{t=l+\frac{k-1}{2}}^{\infty}(\mu^{2})^{t} \frac{\sin(t \alpha)}{t} \Big]
	\notag
	\\
	&=\frac{2}{\alpha}(\mu^{2})^{1-l}\Big[ S_{2l-2+k}(\mu, \frac{\alpha}{2}) - \frac{1}{2}S_{l+\frac{k-1}{2}}(\mu^2, \alpha)\Big],
	\\
	&\sum_{q=l-1+k}\mu^{2q-2l+2-k} \frac{\sin(q\alpha)}{q\alpha}
	=\frac{1}{\alpha}(\mu^{2})^{1-l-k/2} S_{l-1+k}(\mu^{2}, \alpha),
	\\
	&\sum_{q=l-1}^{\infty} \mu^{2q-2l+2+k}\frac{\sin[(2q+k)\frac{\alpha}{2}]}{(2q+k)\frac{\alpha}{2}}
	=\frac{2}{\alpha}(\mu^{2})^{1-l}\sum_{s=2l-2+k, s \text{ odd}} \mu^{s} \frac{\sin(s \frac{\alpha}{2})}{s}
	\notag
	\\
	&=\frac{2}{\alpha}(\mu^{2})^{1-l}\Big[ \sum_{s=2l-2+k} \mu^{s} \frac{\sin(s \frac{\alpha}{2})}{s} 
	- \sum_{s=2l-1+k, s \text{ even}} \mu^{s} \frac{\sin(s \frac{\alpha}{2})}{s} \Big]
	\notag
	\\
	&=\frac{2}{\alpha}(\mu^{2})^{1-l}\Big[ S_{2l-2+k}(\mu, \frac{\alpha}{2}) - \frac{1}{2}\sum_{t=l+\frac{k-1}{2}} (\mu^{2})^{t} \frac{\sin(t \alpha)}{t} \Big]
	\notag
	\\
	&=\frac{2}{\alpha}(\mu^{2})^{1-l}\Big[ S_{2l-2+k}(\mu, \frac{\alpha}{2}) - \frac{1}{2} S_{l+\frac{k-1}{2}}(\mu^{2}, \alpha) \Big],
	\\
	&\sum_{q=l-1}\mu^{2q-2l+2+k} \frac{\sin(q\alpha)}{q\alpha}=\frac{1}{\alpha}(\mu^{2})^{1-l+k/2} S_{l-1}(\mu^{2}, \alpha).
	\label{eq:sumsoddk}
\end{align}
Inserting these equations into Eq.~\eqref{eq:timeevolutionfourier3} we obtain Eq.~\eqref{eq:timeevolutionfourier5}.

\section{Oscillating geometric series\label{app:C}}
We are going to evaluate the sum in Eq.~\eqref{eq:fmu1}. Differentiation with respect to $\mu$ yields
\begin{align}
	\frac{\partial}{\partial \mu} S_{1}(\mu, \alpha)=\sum_{q=1}^{\infty} q \mu^{q-1}\frac{\sin(q\alpha)}{q}=\frac{1}{\mu}\sum_{q=1}^{\infty}\mu^{q} \sin(q\alpha).
	\label{eq:fmuprime1}
\end{align}
Hence we find
\begin{align}
	&\mu \frac{\partial}{\partial \mu} S_{1}(\mu)=\frac{1}{2i} \sum_{q=1}^{\infty} \mu^{q} [\exp(iq\alpha)-\exp(-iq\alpha)]
	\notag
	\\
	&=\frac{1}{2i} \sum_{q=1}^{\infty} \Big(\exp[q(\ln(\mu) +i\alpha)] - \exp[q(\ln(\mu)-i\alpha)] \Big)
	\notag
	\\
	&=\frac{1}{2i} \bigg(  \frac{\exp[(\ln(\mu)+i\alpha)]}{1-\exp[\ln(\mu)+i\alpha]} -  \frac{\exp[(\ln(\mu)-i\alpha)]}{1-\exp[\ln(\mu)-i\alpha]} \bigg)
	\notag
	\\
	&= \frac{1}{2i} \bigg(  \frac{\mu^{k}\exp(i\alpha)}{1-\mu\exp(i\alpha)} -  \frac{\mu^{k}\exp(-i\alpha)}{1-\mu\exp(-i\alpha)}    \bigg)
	\notag
	\\
	&= \frac{\mu}{2i} \big[ (1-\mu\exp[-i\alpha])\exp[i\alpha] 
	- (1-\exp[i\alpha])\exp[-i\alpha]\big] 
	\notag
	\\
	& \phantom{=}/[1+\mu^{2}-2\mu(\exp[i\alpha]+\exp[-i\alpha])]
	\notag
	\\
	&= \frac{\mu }{2i}\big[ \exp(i\alpha)-\exp(-i\alpha) \big]
	/[1+\mu^{2}-2\mu \cos\alpha]
	=\mu\bigg[ \frac{\sin\alpha }{1+\mu^{2}-2\mu\cos\alpha} \bigg].
	\label{eq:fmuprimeovermu}
\end{align}
Thus we have
\begin{align}
	\frac{\partial}{\partial \mu} S_{1}(\mu, \alpha)= \frac{\sin\alpha }{1+\mu^{2}-2\mu\cos\alpha}.
	\label{eq:fmuprime2}
\end{align}
Integration yields
\begin{align}
	S_{1}(\mu, \alpha)=\sin\alpha \int_{0}^{\mu}\diff \lambda \frac{1}{1+\lambda^{2}-2\lambda\cos\alpha}.
	\label{eq:fmu3}
\end{align}
We use the decomposition into partial fractions
\begin{align}
	\frac{1}{1+\lambda^2-2\lambda\cos\alpha} &= 
	\frac{1}{2 i \sin \alpha}\bigg[ \frac{1}{\lambda - \cos \alpha - i \sin \alpha } - \frac{1}{\lambda - \cos \alpha + i \sin \alpha }  \bigg]
	\notag
	\\
	&= \frac{1}{2 i \sin \alpha} \Big[ \frac{1}{\lambda-A} - \frac{1}{\lambda-B} \Big],
	\label{eq:partialfractions}
\end{align}
where we introduced the abbreviations 
\begin{align}
	A= \cos \alpha + i \sin \alpha, 
	\qquad
	B= \cos \alpha - i \sin \alpha.
	\label{eq:abbrevab}
\end{align}
Inserting Eq.~\eqref{eq:partialfractions} into Eq.~\eqref{eq:fmu3} we obtain
\begin{align}
	&S_{1}(\mu, \alpha)= \frac{1}{2i} \int_{0}^{\mu} \diff \lambda \bigg[\frac{1}{\lambda-A} - \frac{1}{\lambda-B} \bigg]
	= \frac{1}{2i} \bigg[\int_{-A}^{\mu-A} \diff \lambda' \frac{1}{\lambda'} -\int_{-B}^{\mu-B}\diff \lambda' \frac{1}{\lambda'} \bigg]
	\notag
	\\
	&=\frac{1}{2i} \Big(\ln\frac{\mu-A}{-A}-\ln\frac{\mu-B}{-B}  \Big)
	=\frac{1}{2i} \Big(\ln\frac{(\mu-A)B}{(\mu-B)A} \Big)
	\notag
	\\
	&=\frac{1}{2i}  \Big[\ln\frac{(\mu-\cos\alpha -i \sin\alpha)(\cos\alpha-i\sin\alpha)}{(\mu-\cos\alpha+i\sin\alpha)(\cos\alpha+i\sin\alpha)}\Big]
	=\frac{1}{2i} \Big[\ln\frac{1-\mu\cos\alpha + i \mu\sin\alpha}{1-\mu\cos\alpha - i\mu\sin\alpha}\Big]
	\notag
	\\
	&=\frac{1}{2i} \bigg\{ \ln\Big[\frac{(1-\mu\cos\alpha)^2 -\mu^2\sin^2\alpha}{ (1-\mu\cos\alpha)^2 +\mu^2\sin^2\alpha} 
	+i \frac{2(1-\mu\cos\alpha)\mu\sin\alpha}{ (1-\mu\cos\alpha)^2 +\mu^2\sin^2\alpha}\Big]\bigg\}
	\notag
	\\
	&=\frac{1}{2i} \bigg\{ \ln 1 + i \arg\Big[\frac{(1-\mu\cos\alpha)^2 -\mu^2\sin^2\alpha}{ (1-\mu\cos\alpha)^2 +\mu^2\sin^2\alpha} 
	+i \frac{2(1-\mu\cos\alpha)\mu\sin\alpha}{ (1-\mu\cos\alpha)^2 +\mu^2\sin^2\alpha}\Big]\bigg\}	\label{eq:fmu4}
	\notag
	\\
	&=\frac{1}{2}\arctan(y,x),
\end{align}
where $x, y$ are given by Eqs.~\eqref{eq:xy1}, \eqref{eq:xy2}.
Obviously for $k>1$ the sum $S_k$ is obtained as
\begin{align}
	S_{k}(\mu, \alpha) = S_{1}(\mu, \alpha) - \sum_{q=1}^{q=k-1} \mu^{q} \frac{\sin(q\alpha)}{q}
	\label{eq:resultfk}
\end{align}
resulting in Eq.~\eqref{eq:fmu2}.
\end{appendix}
\bibliographystyle{iopart-num}
\bibliography{literatur.bib}
\end{document}